# Breaking the Boundaries of the Goldschmidt Tolerance Factor with Ethylammonium Lead Iodide Perovskite Nanocrystals

C. Meric Guvenc[1,2], Stefano Toso[2], Yurii P. Ivanov[3], Gabriele Saleh[2], Sinan Balci[4], Giorgio Divitini[3], Liberato Manna[1]*

[1] Department of Materials Science and Engineering, İzmir Institute of Technology, 35433 Urla, İzmir, Turkey

[2] Nanochemistry, Istituto Italiano di Tecnologia, Via Morego 30, Genova, Italy

[3] Electron Spectroscopy and Nanoscopy, Istituto Italiano di Tecnologia, Via Morego 30, Genova, Italy

[4] Department of Photonics, İzmir Institute of Technology, 35433 Urla, İzmir, Turkey

**ABSTRACT**

We report the synthesis of ethylammonium lead iodide (EAPbI$_3$) colloidal nanocrystals as another member of the lead halide perovskites family. The insertion of an unusually large A-cation (274 pm in diameter) in the perovskite structure, hitherto considered unlikely due to the unfavorable Goldschmidt tolerance factor, results in a significantly larger lattice parameter compared to the Cs-, methylammonium- and formamidinium-based lead halide perovskite homologues. As a consequence, EAPbI$_3$ nanocrystals are highly unstable, evolving to a non-perovskite δ-EAPbI$_3$ polymorph within one day. Also, EAPbI$_3$ nanocrystals are very sensitive to electron irradiation and quickly degrade to PbI$_2$ upon exposure to the electron beam, following a mechanism similar to that of other hybrid lead iodide perovskites (although degradation can be reduced by partially replacing the EA$^+$ ions with Cs$^+$ ions). Interestingly, in some cases during this degradation the formation of an epitaxial interface between (EA$_x$Cs$_{1-x}$)PbI$_3$ and PbI$_2$ is observed. The photoluminescence emission of the EAPbI$_3$ perovskite nanocrystals, albeit being characterized by a low quantum yield (~1%), can be tuned in the 664-690 nm range by regulating their size during the synthesis. The emission efficiency can be improved upon partial alloying at the A site with Cs$^+$ or formamidinium cations. Furthermore, the morphology of the EAPbI$_3$ nanocrystals can be chosen to be either nanocube or nanoplatelet, depending on the synthesis conditions.

**INTRODUCTION**

Lead halide perovskites are a family of direct-gap semiconductors widely investigated as low-cost and high efficiency materials for light emission and harvesting applications.[1,2] For the latter applications, the most promising compositions are the iodine-based $APbI_3$ perovskites, where the *A* site can be occupied by a variety of organic or inorganic monovalent cations. Despite offering high carrier mobility and a nearly ideal gap for solar cells, real-world applications of iodine-based perovskites are hindered by their intrinsic lability, which prompted research into improving the stability of these materials via either cation- or halide-alloying.[1–4] Different $A^+$ cations have been investigated in the attempt to modulate the stability and the properties of lead-iodide perovskites by leveraging the size of cations to influence the Goldschmidt tolerance factor.[3–5] The $APbI_3$ perovskite structure can form with the $A^+$ cation being $Cs^+$ (ionic radius = 188 pm) [6], methylammonium ($MA^+$ = 217)[7], formamidinium ($FA^+$ = 253 pm)[8], and the recently reported aziridinium ($AZ^+$ = 227 pm).[9,10] All these $APbI_3$ perovskite phases are generally unstable under ambient conditions and tend to transform into non-perovskite polymorphs or to degrade (for example to $PbI_2$ and MAI in the case of $MAPbI_3$) over time.[12,15] While some of these phases, like $CsPbI_3$ and $FAPbI_3$, are reasonably durable, the use of significantly smaller or larger cations than $MA^+$ results in more labile phases.[8] Indeed, recent synthesis attempts with dimethylammonium (272 pm), guanidinium (279 pm), and acetamidinium (284 pm) have resulted in the formation of Ruddlesden-Popper phases instead of the perovskite one.[12-14]

Materials that are unattainable in bulk form can however be at times synthesized as nanocrystals, even if only transiently, by exploiting the relaxed structural constraints guaranteed by the finite size of the crystalline domains.[15] Here, we demonstrate the colloidal synthesis of ethylammonium based lead iodide ($EAPbI_3$) perovskite NCs, which display a remarkably high lattice parameter (6.43 Å), and a Goldschmidt factor (1.03) that is outside the boundaries generally considered tolerable for a perovskite phase. Like other lead-halide perovskites, EAPbI$_3$ NCs adopt a cuboidal morphology and display photoluminescence in the 664 - 690 nm range depending on the NCs size. As expected, these EAPbI$_3$ NCs were found to be rather unstable, with their recrystallization to a non-perovskite polymorph starting within ~2h and generally becoming complete within one day, similar to what observed for other APbI$_3$ perovskite phases.[8,11] The low quantum yield values of these EAPbI$_3$ NCs compared to other APbI$_3$ perovskite NCs[11] are likely a consequence of such lability, which might foster the formation of defects and hence trap states. The EAPbI$_3$ perovskite NCs are also extremely sensitive to an electron beam and can degrade to PbI$_2$ even at small irradiation doses, making their characterization challenging. This stability issue can be mitigated by partially replacing EA$^+$ with Cs$^+$. Remarkably, this replacement also leads to the formation of (EA$_x$Cs$_{1-x}$)PbI$_3$ heterostructures, similar to those found for FAPbI$_3$ thin films.[16,17] Partial alloying of EA$^+$ ions with Cs$^+$ or FA cations increases the PL quantum yield of the NCs and induces a spectral red shift, compatible with a band gap narrowing. Notably, the EAPbI$_3$ NCs become more stable when prepared in the form of ultrathin, highly confined nanoplatelets, an effect that is likely due to the relaxation of structural constraints due to their finite thickness, which allows the lattice to cope with the distortions induced by the large size of the EA$^+$ cation.

**RESULTS and DISCUSSION**

**EAPbI$_3$ synthesis and general characterization.** The EAPbI$_3$ NCs were synthesized at room temperature by reacting a PbI$_2$ solution with EA-oleate. In short, PbI$_2$ was dissolved in a mixture of oleylamine and oleic acid in the presence of oleylammonium iodide at 140 °C. Upon cooling, toluene was added to prevent a gel formation at room temperature. Thereafter, EAPbI$_3$ NCs were synthesized by injecting EA-oleate and oleic acid in the PbI$_2$ solution at room temperature. Alternatively, PbI$_2$ could be dissolved in trioctlyphosphine oxide (TOPO) at 140 °C,[18] It should be noted that TOPO is solid at room temperature, and melts only above 70 °C. This limitation can be circumvented by adding toluene to prevent solidification upon cooling the reaction to room temperature. Similarly, to synthesize EAPbI$_3$ perovskite NCs oleylammonium iodide, oleylamine, and oleic acid should be added to the TOPO-PbI$_2$ solution (see the Methods section for detailed synthetic protocols). The size of the colloidal EAPbI$_3$ perovskite NCs could be tuned between 7.1 nm and 17.5 nm depending on the amount of oleic acid added in the synthesis, resulting in tunable photoluminescence (PL) in the 664 – 690 nm range (Figure 1a-c and Figure S1). It should be noted that the EAPbI$_3$ perovskite NCs had weak PL (quantum yield ~1%) and multiexponential PL decay with 1/e lifetime of 37 ns (see Figure S2). We note that EAPbI$_3$ NCs obtained by the two methods described above had comparable optical, structural, and morphological properties (see Figure S3). Low-magnification transmission electron microscopy (TEM) images of EAPbI$_3$ NCs evidenced a generally irregular morphology for the NCs, although many of them had cuboidal shape, similar to that of more conventional lead halide perovskite NCs (Figure 1d). High-angle annular dark-field scanning TEM (HAADF-STEM) imaging supported the assignment of a perovskite crystal structure, with the Fourier transform of the lattice displaying a 4-fold symmetry compatible with a perovskite lattice (Figure 1e). In agreement with electron microscopy, the XRD pattern of the EAPbI$_3$ NCs was compatible with a pseudocubic perovskite structure with a lattice constant of 6.43 Å (Figure 1e-g). This makes EAPbI$_3$ the lead halide perovskite with the largest lattice constant reported to date, despite a Goldschmidt tolerance factor of 1.03, well outside the expected stability range (Figure 1h).[8,12,13]

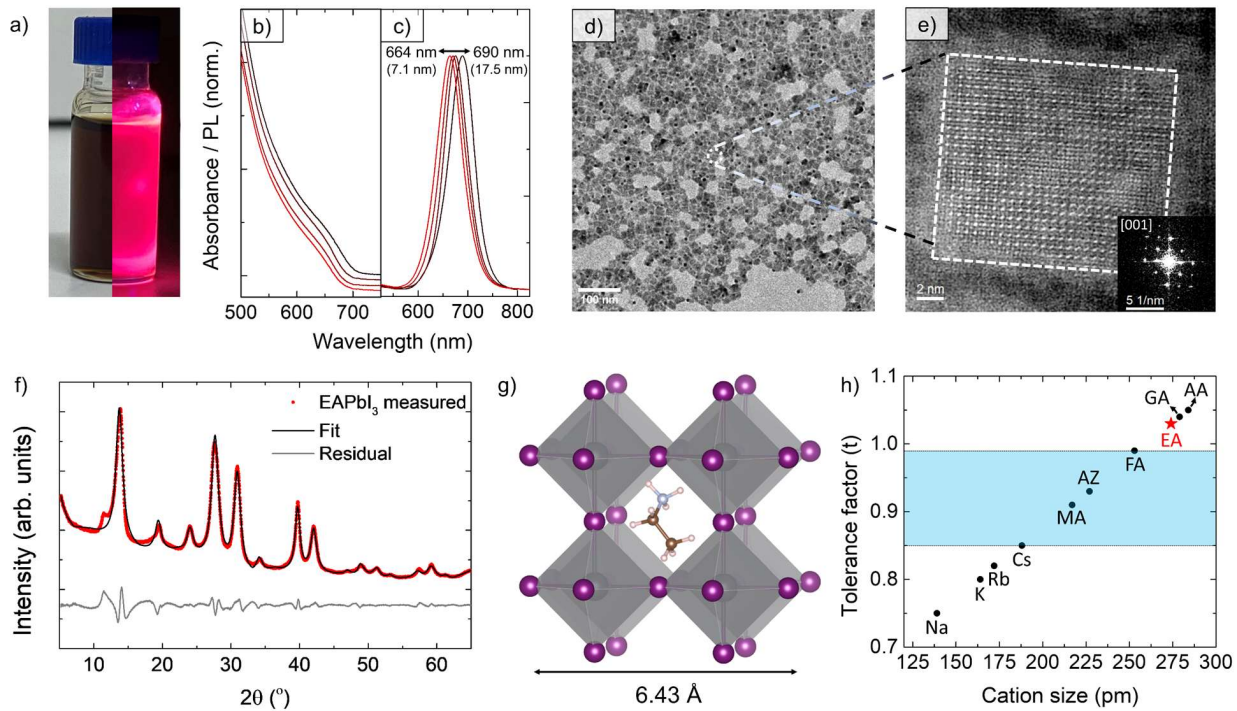

**Figure 1. Optical and structural features of EAPbI₃ NCs.** (a) A colloidal NCs suspension under indoor and UV illumination. (b) Optical absorption and (c) photoluminescence spectra of EAPbI$_3$ NCs of different sizes. (d) Bright field TEM image of EAPbI$_3$ NCs. (e) HAADF HR-STEM image of an individual NC. Inset: Fourier transform of the lattice-resolved image, compatible with a pseudocubic perovskite structure seen along its [001] zone axis. Note that the sample was alloyed with ~12% Cs$^+$ to enhance the stability under the electron beam. (f) X-ray diffraction pattern of EAPbI$_3$ NCs, fitted assuming an R-3c distorted perovskite structure (Le Bail method). (g) Representation of the pseudocubic EAPbI$_3$ crystal structure, with the ethylammonium cation positioned at the center of the cage formed by six [PbI$_6$]$^{4-}$ octahedra. (h) Goldschmidt tolerance factor for APbI$_3$ perovskites with different *A* cations (MA = methylammonium,[7] AZ = aziridinium,[9] FA = formamidinium,[8] GA = guanidinium[13,19] and, AA = acetamidinium[13]). Lead-iodide perovskites obtained experimentally to date are enclosed by the area shown in blue.

An in-depth analysis of the XRD pattern highlighted small discrepancies between the position of the peaks and the cubic indexing. While the ideal perovskite structure provides a decent fit to the XRD profile (see Figure S4), adopting a lower-symmetry space group (for example R-3c in Figure 1e) allows to capture some of these discrepancies. This indicates that the structure of EAPI$_3$ must deviate from that of the ideal perovskite by a mild distortion, similar to what observed for other lead halide perovskites, such as CsPbI$_3$.[20] We note that such deviation is quite minor, and the intrinsic peak broadening due to the nanometric size of crystallites unfortunately prevented us from identifying the correct space group. Our choice of fitting with the R-3c group is only meant to capture the effects of lattice distortions on the XRD profile, but should not be considered a definitive space group attribution to EAPI3. We also note that shoulder at a 2θ value of 11.5° is likely too intense and far from the peak center to be explained by a lattice distortion, and we therefore attribute it to an unidentified byproduct, albeit in small amounts.

In the attempt to further investigate such distortions and gain insights into the electronic structure of EAPbI$_3$ we resorted to density functional theory (DFT) calculations. First, we exploited simulated annealing molecular dynamics to generate 11 low-energy configurations that differed by the position of EA cations, and consequently by the degree of distortion of the Pb-I scaffold (representative examples shown in Figure S5). All of the obtained configurations lie within a fairly narrow energy range of 55 meV/formula unit (f.u.), 36% of them being within 26 meV/f.u. (that is, k$_B$T at 298 K, see Figure S5). Given that there is a vast number of possible ways in which EA cations can be arranged, the small energy difference among the configurations sampled by DFT suggests that many configurations will be populated at room temperature. Within this scenario, we interpret the pseudocubic structure of EAPbI$_3$ as a dynamic average of many local configurations, similarly to what has been reported for methylammonium lead halide perovskites.[20] Such dynamic variability likely has a wide impact on the band gap of EAPbI$_3$, with up to 0.44 eV difference among the configurations inspected (Figure S6). This possibly contributes to the broadening of the excitonic features seen in Figure 1b. Various works on other lead-iodide perovskite phases have reported that the degree of distortion of the Pb-X sublattice can be adopted, in principle, to predict the band gap.[14,21,22] However, we did not find significant correlations (*i.e.* $R^2 \leq 0.32$) between the band gap and any of the tested measures of distortion, including those commonly adopted in the literature (Figure S6). Finally, we note that the electronic structure of EAPbI$_3$ is typical of a lead halide perovskite phase, with valence and conduction bands formed by Pb(6s)-I(5p) and Pb(6p)-I(5p) antibonding orbitals, respectively (Figure S7).[21,24,25]

**2. Instability and transformations of EAPbI$_3$.** As mentioned above, the large ionic radius of EA (274 pm) makes EAPbI$_3$ NCs quite unstable, possibly more than previously reported APbI$_3$ phases (A = Cs$^+$, MA, FA, AZ, see Figure 1g).[16,26] Indeed, the EAPbI$_3$ NCs spontaneously converted to a non-perovskite polymorph, here denoted as δ-EAPbI$_3$ (Figure S8) in analogy with the nomenclature adopted for CsPbI$_3$.[8,11,27,28] Time-dependent XRD analysis (Figure S9) showed that the transformation starts about 2 h after the sample was drop-cast in ambient conditions, and is complete after ~1 day. Similarly, colloidal solutions of EAPbI$_3$ perovskite NCs in toluene were transformed to δ-EAPbI$_3$ within approximately one day at ambient conditions. In some cases, other degradation products were observed in drop cast films, which we interpret as the formation of layered lead halide phases where either EA or excess oleylammonium served as a spacing cation (Figure S10). Notably, transformations in halide perovskite NCs are also easily triggered by exposure to an electron beam in a TEM, which indeed in the present case induced the transformation of EAPbI$_3$ into PbI$_2$, similar to what was reported by Rothmann et al. for FAPbI$_3$.[16,17] Pristine EAPbI$_3$ NCs were converted so quickly that the transformation could not be followed, and only the product PbI$_2$ nanoparticles could be imaged. To slow down such degradation, we therefore replaced ~12% of the EA cations with Cs$^+$ cations via post-synthesis cation exchange to stabilize the NCs during the acquisition of the HAADF-STEM images (Figure 1a-c, see also the Methods section). The concentration of Cs$^+$ was estimated from XRD by using the Vegard's law[29], as discussed in the next section.

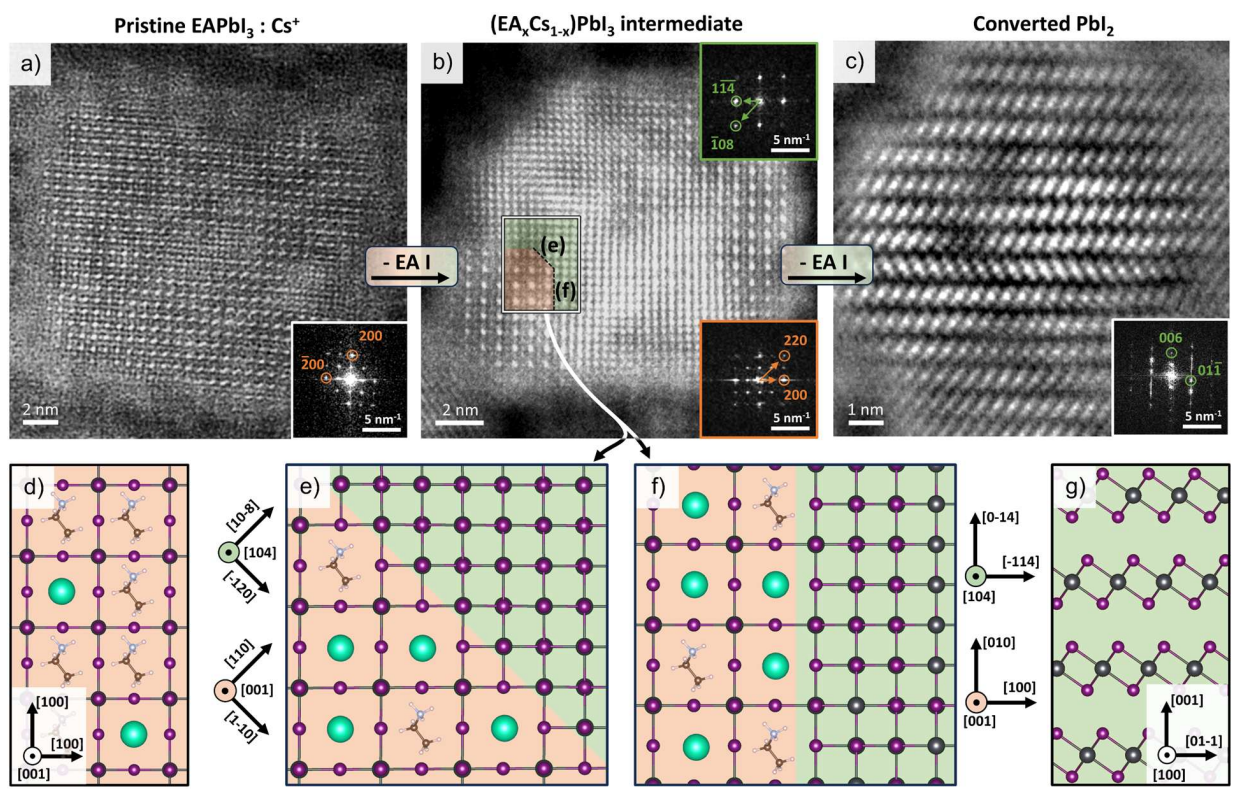

**Figure 2. EAPbI$_3$ → PbI$_2$ degradation mechanism under the electron beam.** (a) HAADF-STEM images of a pristine EAPbI$_3$:Cs$^+$ NC. (b) Partially degraded (EA$_x$Cs$_{1-x}$)PbI$_3$/PbI$_2$ heterostructure, formed as an intermediate while EAI leaves the NC and Cs$^+$ concentrates in the remaining pristine perovskite domains. (c) Fully transformed PbI$_2$ NC, still reminiscent of the initial cuboidal morphology of the perovskite NC. (d-g) Structural models of EAPbI$_3$ (d), PbI$_2$ (g) and of the (EA$_x$Cs$_{1-x}$)PbI$_3$/PbI$_2$ epitaxial interface in two different directions (e,f). The interface reported in panel (f) is structurally analogue to that proposed for FAPbI$_3$/PbI$_2$.[17] The co-presence

of Cs$^+$ and EA$^+$ cations represents the alloyed nature of the perovskite domain (see Figure S12 for further discussion).

Interestingly, the slower degradation of such Cs$^+$-doped particles (Figure 2b) allowed us to gain insight into the EAPbI$_3$ → PbI$_2$ transformation mechanism. The reaction proceeds through the formation of intermediate (EA$_x$Cs$_{1-x}$)PbI$_3$/PbI$_2$ heterostructures (Figure 2d) and is favored by a non-trivial epitaxial relation between the perovskite and PbI$_2$, as proposed in Figure 2e-f, which we identified using the Ogre python library for the prediction of epitaxial interfaces (see Figure S11).[30-32] This is in line with prior observations on the reactivity of lead halide perovskite NCs, which proceeds through reaction intermediates where reagent and product share an epitaxial relation.[16] As the reaction proceeds, the Cs$^+$ initially present as a minority cation is expelled from areas transformed into PbI$_2$, and being less volatile than EA$^+$ it eventually accumulates into (EA$_x$Cs$_{1-x}$)PbI$_3$ domains, which are the last ones to be converted (see Figure S12 for additional analyses of the (EA$_x$Cs$_{1-x}$)PbI$_3$/PbI$_2$ interface).

**Alloys with FA and Cs.** The stability enhancement achieved by partially replacing EA cations with Cs$^+$ cations prompted us to explore (EA$_x$FA$_{1-x}$)PbI$_3$ and (EA$_x$Cs$_{1-x}$)PbI$_3$ alloys, where a substantial fraction of EA is replaced with smaller FA and Cs$^+$ cations, respectively. As mentioned in the previous section, FA and Cs cations were introduced into the EAPbI$_3$ perovskite structure via post-synthesis cation exchange (See Methods section for details). In both cases, the compositional tuning induced a red shift of both the absorption edge (Figure 3a,c) and the PL (Figure 3b,d), which was significantly more marked for FA$^+$ than for Cs$^+$. Also, the PQLY of the FA$^+$ and Cs$^+$ alloyed EAPbI$_3$ NCs increased (Figure S13). The partial exchange with Cs$^+$ had a more prominent effect on the XRD pattern (Figure 3e,f), where the unit cell contraction is more significant due to the smaller ionic radius of cesium. Based on the Vegard's law[29], we estimated that the maximum exchange ratios reached ~55:45 and ~30:70 for EA:FA and EA:Cs in alloyed EAPbI$_3$ NCs, respectively. (see Table S1 and S2).

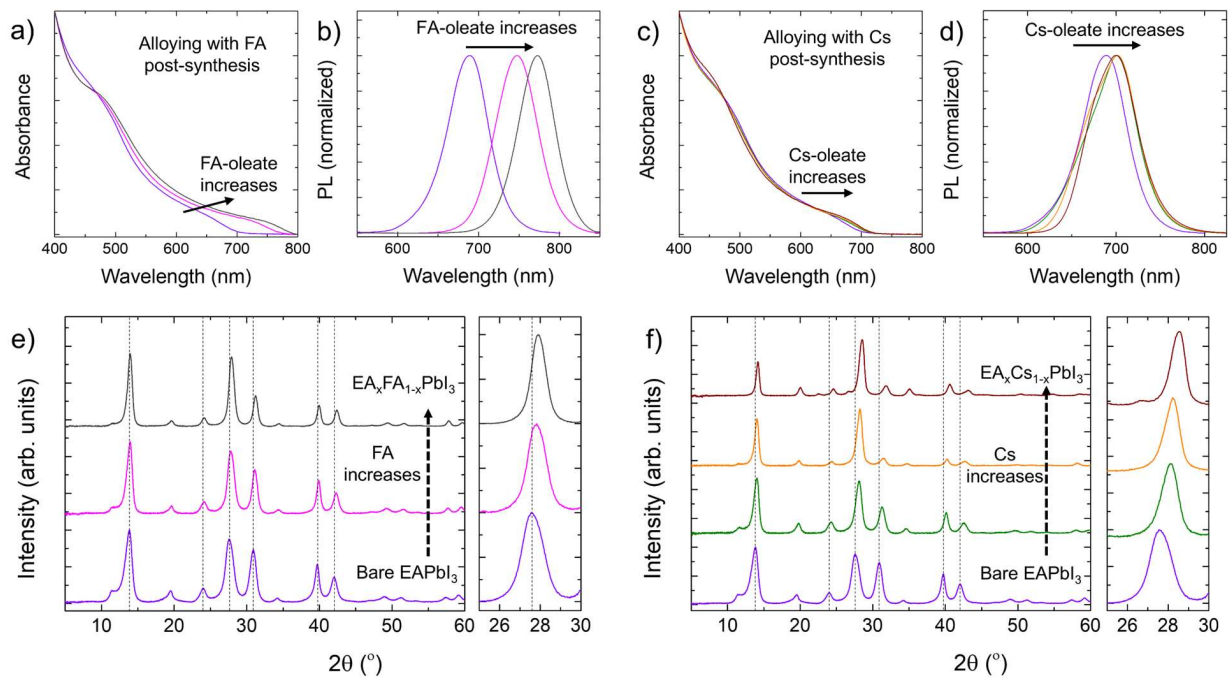

**Figure 3. Alloys with FA and Cs.** Characterization of (EA$_x$FA$_{1-x}$)PbI$_3$ NCs (left) and (EA$_x$Cs$_{1-x}$)PbI$_3$ NCs (right). From left to right and from top to bottom: optical absorption spectra (a,c); PL spectra (b,d); XRD patterns (e,f). The cation alloying ratios are extracted from XRD patterns in panels (e, f) by using Vegard's law.[29] The EA:FA ratios for (EA$_x$FA$_{1-x}$)PbI$_3$ NCs were estimated to be 100:0 (violet), 60:30 (pink), and 55:45 (gray) (e). The EA:Cs ratios in (EA$_x$Cs$_{1-x}$)PbI$_3$ NCs were determined to be approximately 100:0 (violet), 75:25 (green), 60:40 (orange), and 30:70 (red) from bottom to top (f).

The more marked spectral shift induced by FA$^+$ might appear counterintuitive, given that the unit cell actually contracts less than in the Cs$^+$ case. However, a similarly non-linear dependence of the band gap when introducing larger *A* cations was reported for related Ruddlesden-Popper lead-iodide phases,[13] where it was rationalized as the combination of a gap widening due to the stretching of Pb-I bonds plus a gap narrowing due to the reduced octahedra tilting. In this light, we justify the stronger spectral shift induced by $FA^+$ with a shortening of the Pb-I bonds ($EA^+$ = 274 pm, $FA^+$ = 253 pm) accompanied by virtually no octahedra tilting, as both $EAPbI_3$ and $FAPbI_3$ adopt structures close to the ideal cubic archetype. Conversely, the introduction of the much smaller $Cs^+$ cation cannot shorten the Pb-I bonds much further, due to physical limits in the inter-ionic distance ($FA^+$ = 253 pm, $Cs^+$ = 188 pm), but it does cause major deviations from the ideal cubic symmetry. This is supported by extra peaks appearing in the XRD pattern of $(EA_xCs_{1-x})PbI_3$ in the 22-27° 2θ range, that are typical of a heavily distorted orthorhombic perovskite structure. In conclusion, the opposite effect of Pb-I bonds shortening and induced octahedra tilting likely balances out in the case of $Cs^+$ alloying, leaving the spectral properties of $(EA_xCs_{1-x})PbI_3$ alloys almost unaffected.

**$EAPbI_3$ nanoplatelets.** Besides alloying, it is known that perovskites with large $A^+$ cations can be partially stabilized by adopting a thin platelet morphology.[13] The lack of structural constraints in the thin direction of the platelets generates an element of anisotropy extrinsic to the crystal structure of the material,[12,13] and allows it to accommodate distortions that would not be compatible with a more extended crystal of the same material. Such mechanism justifies the stability of Ruddlesden-Popper lead-iodide phases[12] when compared to their 3D-perovskite $APbI_3$ counterparts: here, the Pb-I octahedra that compose their disconnected layers can adopt tilting motifs that would be incompatible with a 3D-connected structure, but allow them to better accommodate the $A^+$ cations. This is reflected in the Pb-Pb distances, which tend to differ sensibly from those measured in the corresponding $APbI_3$ 3D-perovskites and can sometimes become anisotropic in the two directions of the lattice (i.e., in the octahedra plane vs in the stacking direction of layers), which reflects the adoption of octahedral tilting and distortions different from those of 3D perovskites. Analogous effects were recently demonstrated also for colloidal perovskite NPLs.[33-35] Following this direction, we further modified our initial synthesis protocol to induce the formation of EAPbI$_3$ nanoplatelets, which was achieved by gradually reducing the amount of EA-oleate injected (see Methods). This caused a progressive blue-shift of both the optical absorption and PL spectra, concomitant with the formation of a sharp excitonic peak typical of highly confined perovskite NCs (Figure 4a-b). [36-38]

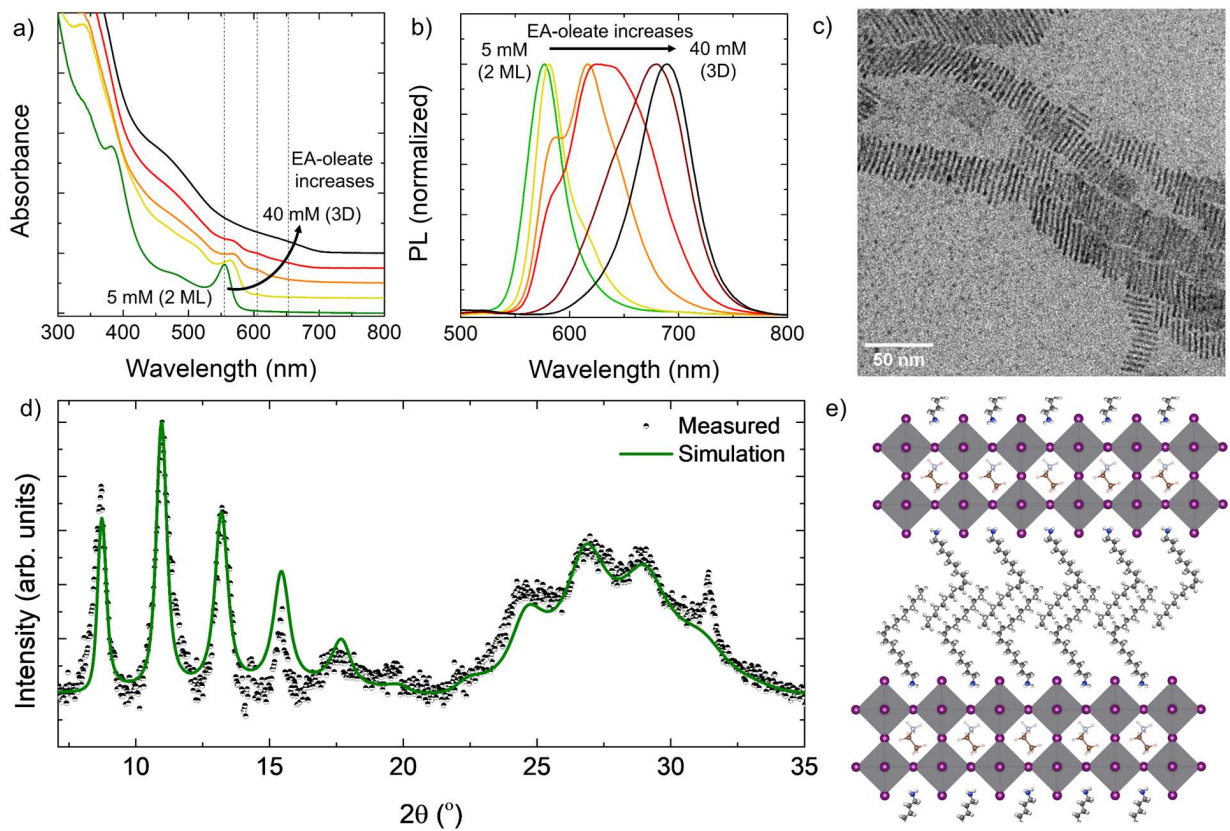

**Figure 4. EAPbI$_3$ nanoplatelets.** (a, b) Optical absorption (a) and PL (b) spectra of EAPbI$_3$ samples obtained with different concentrations of EA-oleate in the synthesis. The samples prepared with low EA-oleate concentration show the onset of a sharp excitonic peak, attributed to EAPbI$_3$ nanoplatelets with a thickness as small as 2 PbI$_6$ octahedra (2ML in short). (c) TEM images of 2ML EAPbI$_3$, highlighting the characteristic face-to-face stacking. (d) XRD pattern of 2ML EAPbI$_3$ analyzed via multilayer diffraction. The characteristic series of periodic peaks is a signature of self-assembly of platelets into flat and ordered stacks.[33] (e) Structure representation of a stack of 2ML EAPbI$_3$ nanoplatelets, constructed according to the structural parameters extracted from the multilayer diffraction fit.

The successful formation of nanoplatelets was independently confirmed by TEM (Figure 4c), from which it could be seen that the particles adopt the characteristic face-to-face stacked assembly. Larger area TEM images of the nanoplatelets are reported in Figure S14. For the most confined platelets, the position of the spectral features (abs = 550 nm, PL = 577 nm) suggests a thickness of two octahedra mono-layers, (2 ML, Figure 4c), which can be gradually increased by adding more EA-oleate during the synthesis. However, this results in a lower level of control over their thickness distribution, as samples containing mostly 3ML nanoplatelets already exhibited a shoulder in their PL spectrum that is compatible with higher thicknesses (Figure 4b). TEM images of the mixed thickness nanoplatelets are reported in Figure S15. An in-depth inspection of the XRD pattern of 2ML EAPbI$_3$ nanoplatelets, performed via multilayer diffraction analysis,[33] revealed major deviations from the crystal structure of EAPbI$_3$ nanocubes. In particular, the Pb-Pb distance measured in the platelets is significantly smaller than what we found in EAPbI$_3$ nanocubes, 6.31 Å in platelets vs 6.43 Å in nanocubes (see Figure S16 and S17 for further discussion). This is indirect proof that the position of the Pb-I octahedra in the platelets must be different from that in nanocubes. Unfortunately, due to the insufficient quality of the diffraction pattern we could not refine the octahedral tilting via multilayer diffraction. However, the shorter Pb-Pb distance in the platelets indicates that these are able to partially relax the tension imposed on the Pb-I bonds by the exceedingly large ionic radius of EA$^+$. Indeed, we observed that EAPbI$_3$ nanoplatelets are significantly more robust than nanocubes of the same material, as little to no sign of degradation was observed even after several days from the synthesis (see Figure S18).

**CONCLUSION**

In this work, we demonstrate the synthesis of ethylammonium lead iodide EAPbI$_3$ perovskite in the form of colloidal NCs. As a material, EAPbI$_3$ perovskite had never been reported previously. The EA cation was considered unable to form a 3D lead-halide perovskite structure due to its large ionic radius (~274 pm), resulting in anunfavorable Goldschmidt tolerance factor (1.03). Notably, EAPbI$_3$ is the lead-iodide perovskite with the largest lattice constant (when compared with NCs of similar size). This makes EAPbI$_3$ an interesting material for validating predictions of the electronic and crystal structure of lead iodide perovskites beyond the boundaries of previously reported phases.

**METHODS**

***Chemicals.*** Lead (II) iodide (PbI$_2$, 99%), Cesium carbonate (Cs$_2$CO$_3$, 99.9%), Formamidine acetate salt (FA-acetate, HN=CHNH$_2$·CH$_3$COOH, 99.9%), Ethylamine solution (2.0 M in tetrahyrofuran), oleylamine (OLAM, 70%), hydroiodic acid (HI, 57 wt. % in water), oleic acid (OA, 90%), Trioctylphosphine oxide (TOPO, 99%) and toluene were all purchased from Sigma-Aldrich except TOPO was purchased from Strem chemicals and used without any further purification.

***Synthesis of OLAM-I.*** 10 mL of OLAM and 1.68 mL of HI were loaded in a 100 mL three neck round bottomed flask. The mixture was heated to 130 °C for 2 hours under the nitrogen flow. Subsequently, the reaction was transferred to a vial while it was hot and then cooled to room temperature. 10 mL of dense OLAM-I solution prepared as described above was diluted with 14 mL OLAM for using experiments. OLAM-I solution gels at room temperature therefore prior to use, OLAM-I precursor heated at 120 °C.

***Synthesis of EA-Oleate.*** 1.25 mL 2 M ethylamine solution in THF, and 2 mL of OA were mixed in a vial for 2 hours at room temperature. Afterwards, 6.75 mL of toluene was added into the vial. Prepared solution stored for further use.

***Synthesis of Cs-Oleate.*** 407 mg of $Cs_2CO_3$, and 2 mL of OA were mixed in a round bottom flask. The mixture was degassed at room temperature for 10 min, and then further degassed for 30 min at 120 °C. Temperature of the reaction vessel was set to 135 °C under the flow of nitrogen gas and kept at the same temperature until a clear solution was obtained. Afterwards, 8 mL of toluene was quickly injected into the reaction flask and the reaction mixture was cooled down to room temperature. Cs-oleate often precipitate at room temperature. Cs-oleate was heated at 100 °C until all precipitate was dissolved prior to use.

***Synthesis of FA-Oleate.*** 260 mg of FA-acetate and 2 mL of OA were loaded in a round bottom flask. The mixture was degassed at room temperature for 10 minutes. After that, the solution was inserted preheated oil bath at 130 °C under the nitrogen atmosphere for 5 minutes. Subsequently, the flask was removed from the oil bath and cooled for 2 minutes at ambient temperature. Later, the solution was dried at 55 °C under the vacuum for 10 minutes. And then, the FA-oleate solution was cooled to room temperature and stored for further use. The FA-oleate solution was preheated to 120 °C prior to use until the solution became clear.

***Synthesis of Pb-precursor with OLAM-I and OA.*** 0.4 mmol $PbI_2$, 500 μL of OLAM-I, and 400 μL of OA were loaded in a glass tube and degassed at 80 °C and subsequently, heated to 140 °C under vigorous stirring until all $PbI_2$ solution became completely clear. Then, 4 mL of toluene was added into the reaction mixture. The precursor solution was cooled down to room temperature and stored for further use.

***Synthesis of Pb-precursor with TOPO.*** 0.4 mmol PbI$_2$, and 400 mg of TOPO were loaded in a glass vial and, heated to 140 °C under vigorous stirring until all PbI$_2$ solution became completely clear. Then, 4 mL of toluene was added into the reaction mixture. The precursor solution was cooled down to room temperature and stored for further use.

***EAPbI$_3$ NC synthesis from Pb-precursor with OLAM-I and OA.*** 1 mL of toluene, 200 µL Pb-precursor, and 300 µL of EA-oleate were added in a vial, and a slightly yellow solution was obtained. After the addition of 60-300 µL of the oleic acid, the reaction immediately started, and the solution color was changed to black. The higher amount of the oleic acid used in the synthesis causes the formation of more quantum-confined NCs. The obtained NC solution was centrifuged at 4000 rpm for 2 minutes, and the precipitate and supernatant separated. Then, the precipitate was dispersed in toluene. Further, the supernatant was centrifuged at 14500 rpm for 10 minutes, then the precipitate dispersed toluene, and the supernatant was discarded.

***EAPbI$_3$ NC synthesis from Pb-precursor with TOPO.*** 500 µL of toluene, 100 µL Pb-precursor (TOPO), 7 µL of OLAM-I solution, and 150 µL of EA-oleate were added in a vial, and a slightly yellow solution was obtained. After the addition of 100-800 µL of the oleic acid, the reaction immediately started, and the solution color was changed to black. The higher amount of the oleic acid used in the synthesis causes the formation of more quantum-confined NCs. The obtained NC solution was centrifuged at 6000 rpm for 1 minutes and precipitate discarded and supernatant centrifuged again at 14500 rpm for 6 minutes, and supernatant discarded and precipitate dispersed in toluene. Sometimes precipitate can only be dispersed with the help of the ultrasonication.

***EAPbI$_3$ NPL synthesis.*** 1 mL of the toluene, 200 µL Pb-precursor (OLAM-I and OA), and 35 µl of EA-oleate mixed, respectively. Then, 200 µL of the oleic acid was added to this solution, and EAPbI$_3$ NPLs were immediately formed. This solution was centrifuged at 6000 rpm for 1 minute, and the precipitate was redispersed in toluene. After this NPL solution was centrifuged at 6000 rpm for 2 minutes, the precipitate was discarded, and the supernatant was taken for further use. For the TOPO Pb-precursor, 500 μL toluene, 100 μL Pb-oleate, 7.5 μL OLAM-I solution, and 17.5 μL EA-oleate mixed, respectively. After 200 μL OA was added, EAPbI$_3$ NPLs were immediately formed. The same centrifuge procedure described above was applied to separate the NPLs.

***Cs or FA alloyed EAPbI$_3$ NC synthesis.*** After the synthesis of the EAPbI$_3$ perovskite NCs. (EA$_x$FA$_{1-x}$)PbI$_3$ NCs 7 μL or 14 μL FA-oleate solution, for (EA$_x$Cs$_{1-x}$)PbI$_3$ NCs 3.5 μL, 7 μL, or 14 μL Cs-oleate solution was added to EAPbI$_3$ crude NC solution. It should be noted that half the amount of FA-oleate and Cs-oleate described above was used to alloy EAPbI$_3$ NCs synthesized using the TOPO route, as the TOPO synthesis contained half the amount of Pb and EA precursors. The cleaning procedure was the same as the EAPbI$_3$ NCs cleaning procedure described above; it only was changed depending on which PbI$_2$-precursor type was used.

*Optical characterization*

UV-Vis absorption spectra were obtained using a Varian Cary 300 UV−Vis absorption spectrophotometer (Agilent). The spectra were collected by diluting 40 μL of the sample in toluene in 3 mL of toluene. Photoluminescence spectra were obtained on a Varian Cary Eclipse Spectrophotometer (Agilent). Time-resolved photoluminescence spectra were obtained using an Edinburgh FLS900 fluorescence spectrophotometer PL quantum yield measurements. PL decay traces were measured with a 508 nm picosecond pulsed laser diode (EPL-510, Edinburgh Instruments). Quantum yield measurements were acquired using a calibrated integrating sphere with λ$_{ex}$ = 350 nm for all measurements (FS-5, Edinburgh Instruments). All solutions were diluted to an optical density of 0.1 - 0.2 at the excitation wavelength to minimize the reabsorption of the fluorophore. Quartz cuvettes with an optical path length of 1 cm were used for all-optical analyses.

*Powder X-ray Diffraction (XRD) Analysis*

XRD patterns were obtained using a PANalytical Empyrean X-ray diffractometer equipped with a 1.8 kW Cu Kα ceramic X-ray tube and a PIXcel3D 2 × 2 area detector operating at 45 kV and 40 mA. The diffraction patterns were collected in the air at room temperature using parallel beam (PB) geometry and symmetric reflection mode. All XRD samples were prepared by drop-casting a concentrated solution on a zero-diffraction quartz wafer.

The Vegard's law [29,39,40] analysis of $A^+$ cation alloys was performed by extracting the pseudocubic lattice parameters of samples via Le Bail fitting, like shown in Figure S3. The same approach was adopted to extract the lattice constants of pure $CsPbI_3$ (6.216 Å) and $FAPbI_3$ (6.346 Å) NCs, which serve as references for the application of Vegard's law. We opted not to adopt published refences because the lattice constant of NCs can be slightly different from that of the corresponding bulk, and to ensure a consistent treatment and error cancellation, if present.

The multilayer diffraction analysis of $EAPbI_3$ nanoplatelets was performed using the code published.[33] Due to the lack of an established bulk structure for $EAPbI_3$, and to the likely disordered position of the $EA^+$ cations, we opted to model its electron density by including in the multilayer model 2 atoms of carbon and 1 of nitrogen at the center of the unit cell. This choice is adequate for a preliminary modelling, as the $EA^+$ contribution to the total electron density is negligible (18 electrons / formula unit, excluding hydrogens) compared to the contribution of heavy atoms ($Pb^{2+}$ + 3 $I^-$, 242 electrons / formula unit). The impact of EA becomes even smaller when considering that the actual stoichiometry of such nanoplatelets is $(OLAM)_2EAPb_2I_7$, as their thickness is just 2 $PbI_6$ octahedra.

*Electron Microscopy*

Bright field TEM images were acquired on a JEOL JEM-1400 microscope equipped with a thermionic gun at an accelerating voltage of 120 kV. The samples were prepared by drop-casting diluted NC suspensions onto 200 mesh carbon-coated copper grids.

High-resolution scanning transmission electron microscopy (STEM) images were acquired on a probe-corrected Thermo Fisher Spectra 30–300 S/TEM operated at 300 kV. Atomic resolution images were acquired on a high-angle annular dark field (HAADF) detector with a current of 30 pA and a beam convergence semiangle of 25 mrad.

*Atomistic simulations*

All DFT simulations were performed with the VASP software[41] within the projector augmented plane wave[42] and adopting the PBE functional[43] For the simulated annealing molecular dynamics (MD) runs, the initial structure was generated by substituting -H with -CH3 in the ground state of methylammonium lead halide perovskite[44] and contained 8 formula units. The temperature was increased to 1000 K in 5000 steps (time step 1 fs, NpT ensemble, Langevin thermostat). Then, the MD was run for further 50000 steps, taking one structure every 5000 steps and cooling it down to T=200 K in 5000 steps (slower cooling -40000 steps- adopted for those structures that would otherwise lose their perovskite structure upon cooling; one of them still cooled down in a non-perovskite structure and was thus discarded). Two additional structures were obtained with milder annealing at T=450 K. All these ab initio MD simulations were performed with gamma-point only reciprocal space sampling and a plane-wave energy cutoff of 320 eV. All the structures obtained were then tightly relaxed ($\Delta E=10^{-6}$) with a 2x2x2 k points sampling and an energy cutoff of 500 eV, adding the Tkatchenko-Sheffler correction[44] to account for dispersion forces, which was shown to give accurate results in methylammonium lead halide perovskites.[46]

## ASSOCIATED CONTENT

**Supporting Information**.

Supplemental figures and tables, $EAPbI_3$ nanoplatelets analysis, $EAPbI_3/PbI_2$ epitaxial interface modeling.

## AUTHOR INFORMATION

**Corresponding Author**

*liberato.manna@iit.it

**Present Addresses**

†Division of Chemical Physics, Lund University, Lund 221 00, Sweden

**Author Contributions**

The manuscript was written through contributions of all authors. All authors have given approval to the final version of the manuscript.

**Funding Sources**

Cetin Meric Guvenc acknowledges the support of TUBITAK (The Scientific and Technological Research Council of Türkiye) within the 2214-A - International Research Fellowship Programme for Ph.D. students. L. M. acknowledges funding from European Research Council through the ERC Advanced Grant NEHA (grant agreement n. 101095974).

## ACKNOWLEDGMENT

We acknowledge the materials characterization facility at Istituto Italiano di Tecnologia providing access to the PANalytical Empyrean X-Ray Diffractometer. We also acknowledge the computing resources and the related technical support used for this work have been provided by CRESCO/ENEAGRID High-Performance Computing infrastructure and its staff. CRESCO/ENEAGRID High-Performance Computing infrastructure is funded by ENEA, the Italian National Agency for New Technologies, Energy and Sustainable Economic Development, and by Italian and European research programs (http://www.cresco.enea.it/english).

*Supporting Information for*

***Breaking the Boundaries of the Goldschmidt Tolerance Factor with Ethylammonium Lead Iodide Perovskite Nanocrystals***

C. Meric Guvenc[1,2], Stefano Toso[2], Yurii P. Ivanov[3], Gabriele Saleh[2], Sinan Balci[4], Giorgio Divitini[3], Liberato Manna[1*]

[1] Department of Materials Science and Engineering, İzmir Institute of Technology, 35433 Urla, İzmir, Turkey

[2] Nanochemistry, Istituto Italiano di Tecnologia, Via Morego 30, Genova, Italy

[3] Electron Spectroscopy and Nanoscopy, Istituto Italiano di Tecnologia, Via Morego 30, Genova, Italy

[4] Department of Photonics, İzmir Institute of Technology, 35433 Urla, İzmir, Turkey

**Contents**

# S1. Supplemental figures and tables

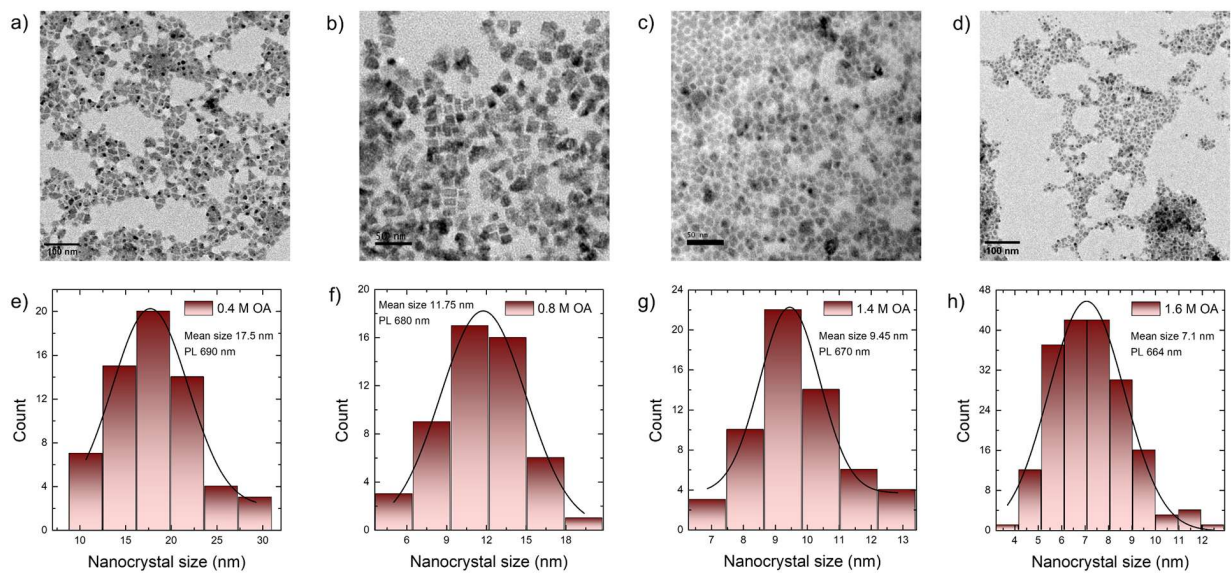

**Figure S1.** (a, b, c, d) TEM images of the EAPbI$_3$ perovskite nanocrystals, which contain 0.4 M, 0.8 M, 1.4 M, and 1.6 M OA in the final solution, and (d, e, f, h) their related particle size distributions, respectively.

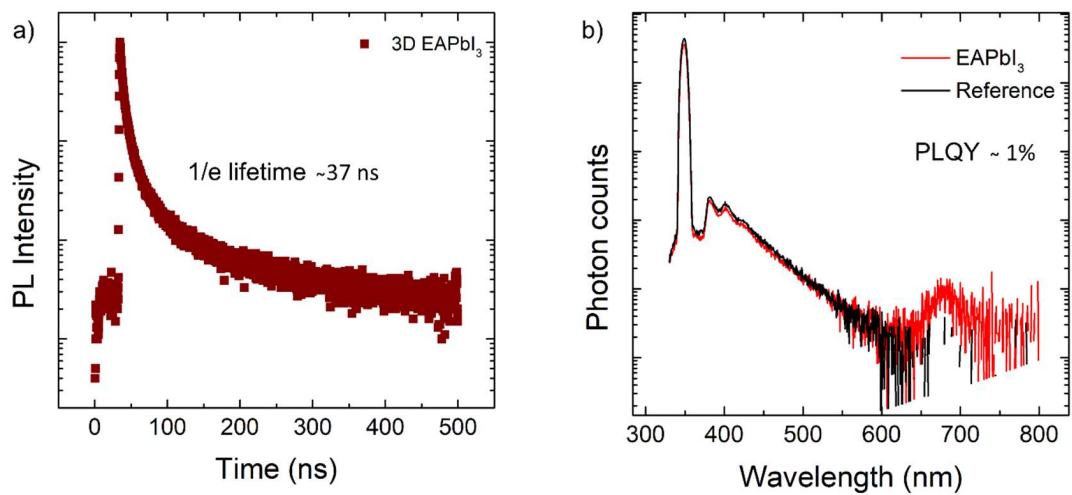

**Figure S2.** (a) PL intensity decay curve and (b) spectra recorded for PLQY measurement of EAPbI$_3$ perovskite nanocrystals.

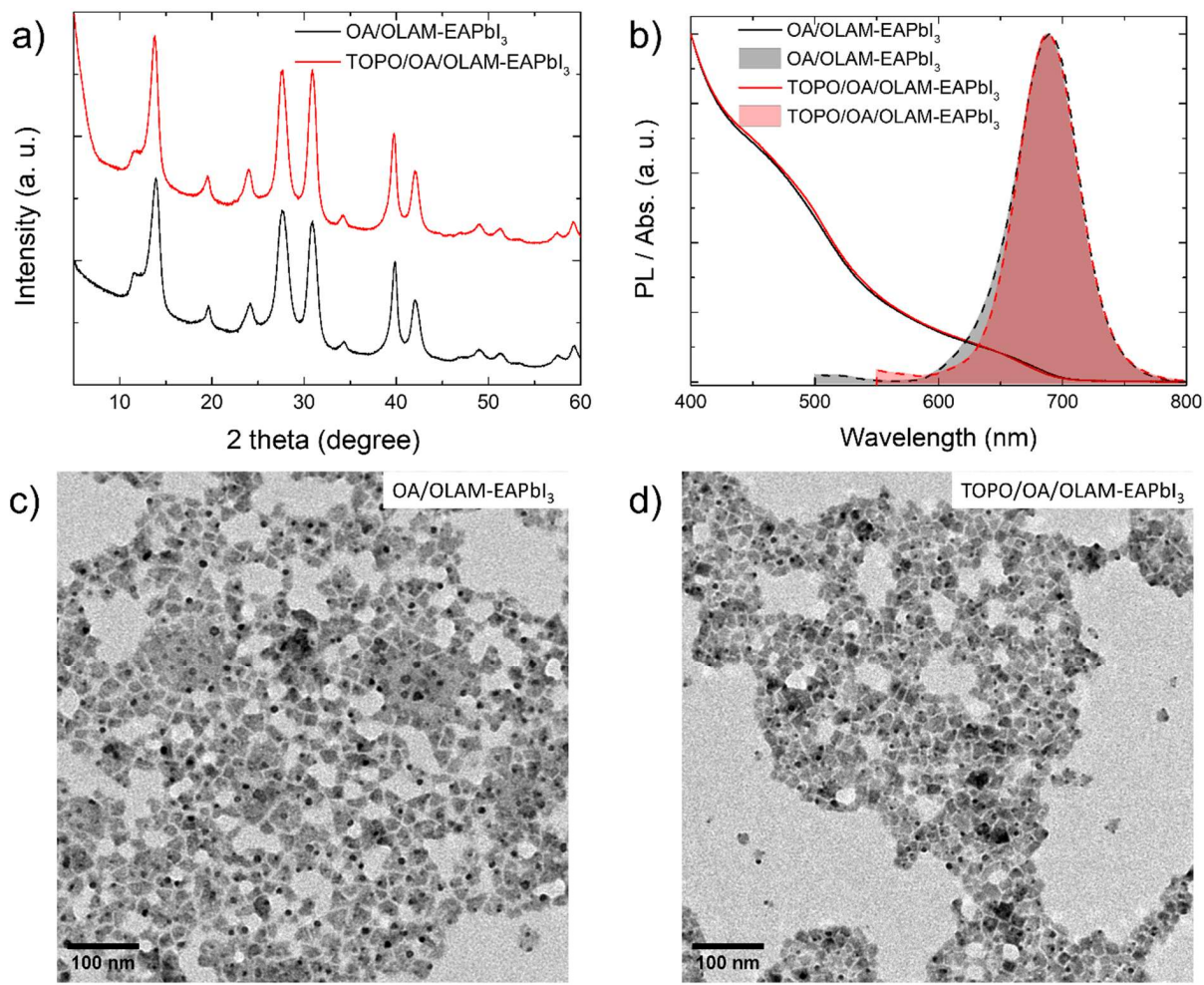

**Figure S3** Comparison of the absorption, PL spectra, TEM images, and XRD patterns of the EAPbI$_3$ NCs synthesized with two methods used in this study. XRD patterns, absorption, PL spectra, and morphology of the EAPbI$_3$ NCs are very similar for both methods.

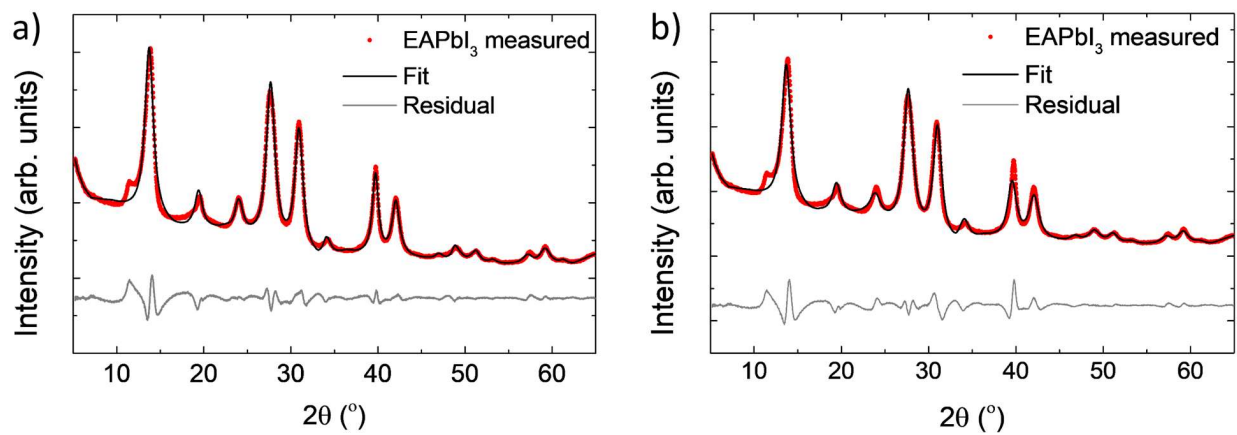

**Figure S4.** Experimental XRD data of the EAPbI$_3$ perovskite fitted with (a) R-3c and (b) Pm-3m prototype.

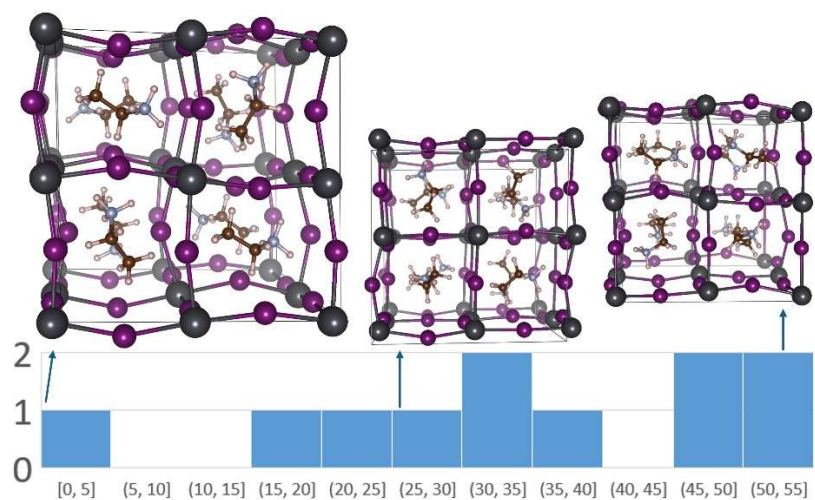

**Figure S5** Histogram showing the energy distribution (meV/formula unit) of the 11 EAPbI$_3$ configurations obtained through simulated annealing. Representative configurations are shown. the larger, left most one is the lowest-energy structure (putative structural ground state) obtained. Color code for atoms as follows. Pb= grey, I=violet, C=brown, N=blue, H=white.

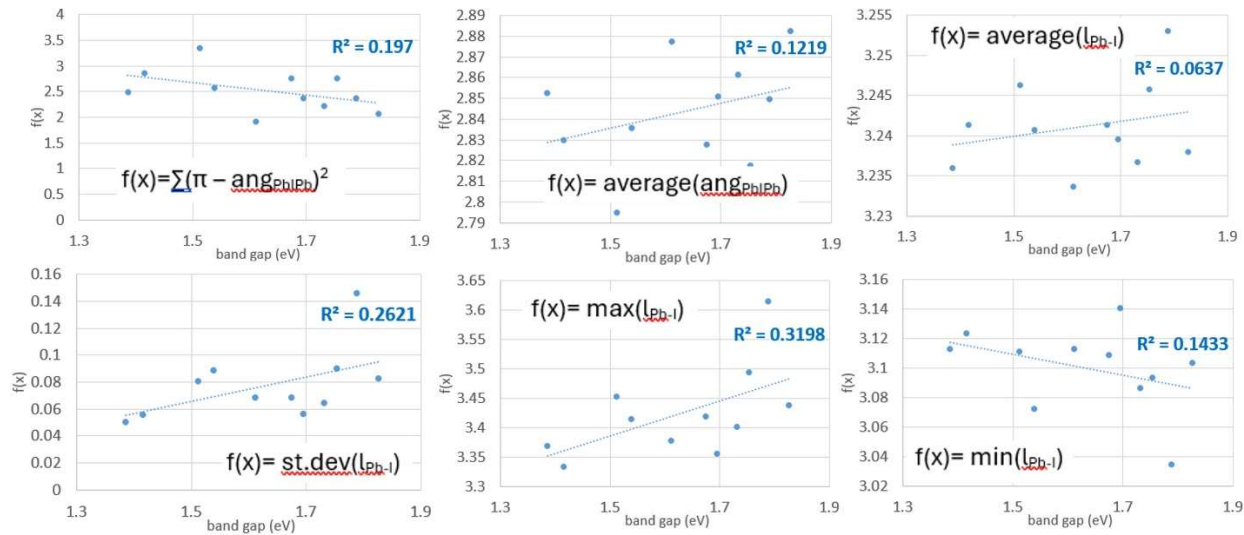

**Figure S6** Plots of the band gap vs various measures of Pb-I sublattice distortion for the 11 studied EAPbI$_3$ configurations. For each plot, the function reported on the vertical axis is shown in black inside the plot. ang$_{PbIPb}$ is the angle formed by Pb with two bonded I atoms, and l$_{Pb-I}$ is the Pb-I bond length. The Pearson correlation coefficient squared ($R^2$) is reported in blue in each plot.

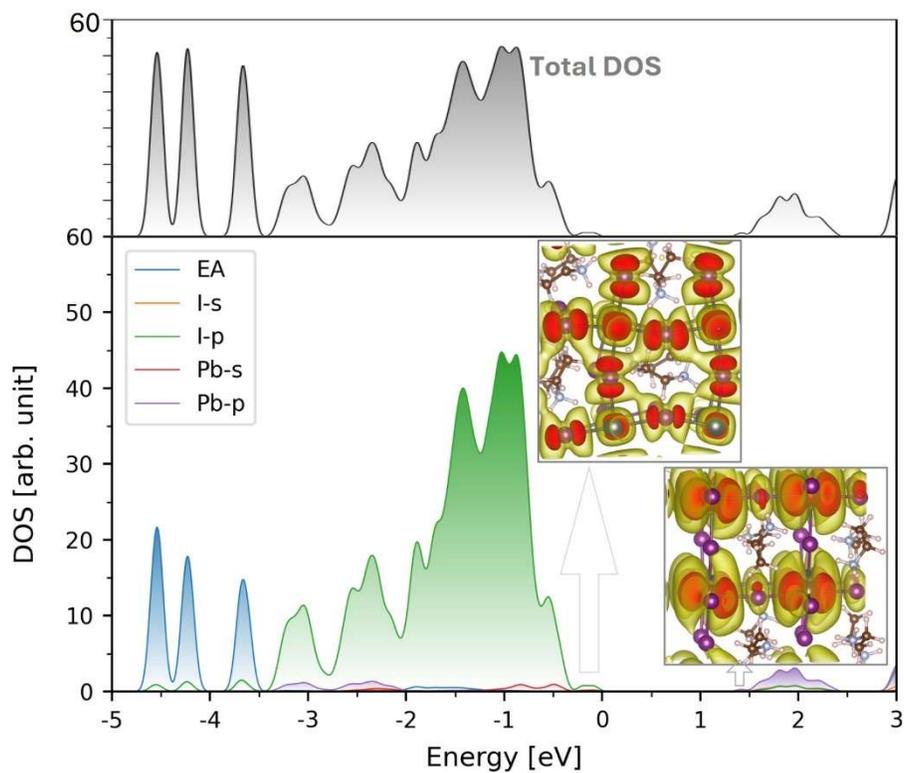

**Figure S7** Density of states (DOS) and partial charge density for EAPbI$_3$ (lowest-energy configuration). The top plot reports the total DOS, while the bottom plot shows the atomic orbital-projected contributions, color legend on the top left. The charge density corresponding to the valence band maximum (from -0.25 eV to Fermi level) and conduction band minimum (from Fermi level to + 1.5 eV) are reported in the insets as yellow and red isosurfaces (isovalues, in VASP units: 5*10$^{-5}$ and 5*10$^{-4}$ for valence and 1*10$^{-5}$ and 1*10$^{-4}$ for conduction). Color code same as in Fig. S5.

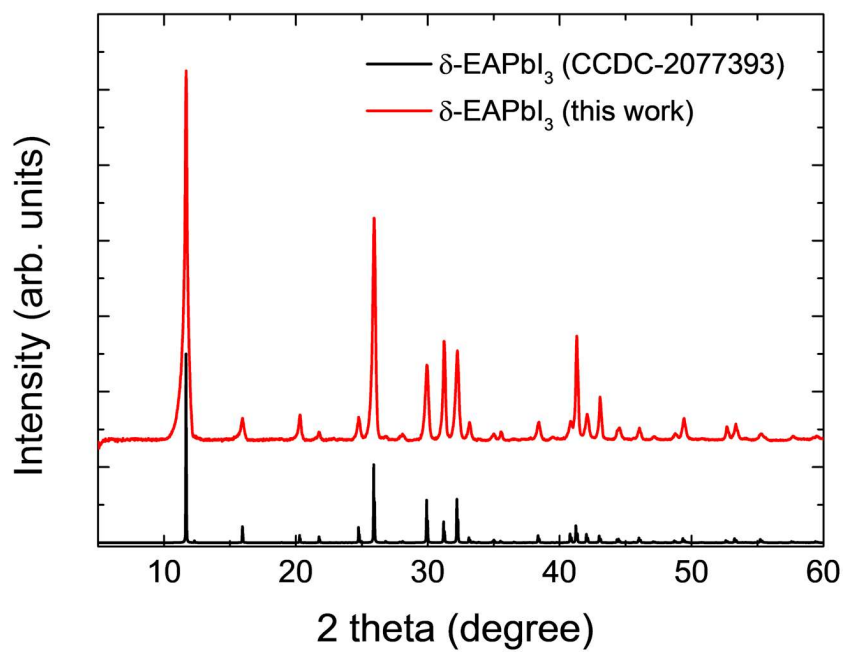

**Figure S8.** Comparison of the non-perovskite δ-EAPbI$_3$ crystals synthesized in this work and reference δ-EAPbI$_3$ (CCDC-2077393) XRD patterns.[1]

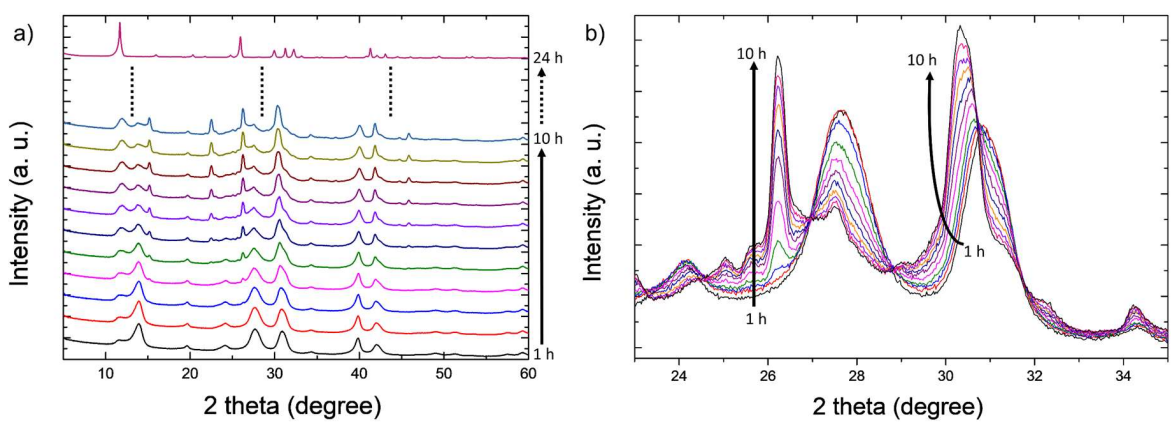

**Figure S9.** (a, b) The time-dependent XRD measurements of the EAPbI$_3$ perovskite nanocrystals. XRD analysis demonstrates that phase transformation starts after 2 hours in the drop-cast films of the EAPbI$_3$ nanocrystals under ambient conditions. After 24 hours, the EAPbI$_3$ perovskite phase

completely transforms to δ-EAPbI₃ phase. (b) It is obvious that peak intensities change even 10 hours later, and phase transformation was not complete.

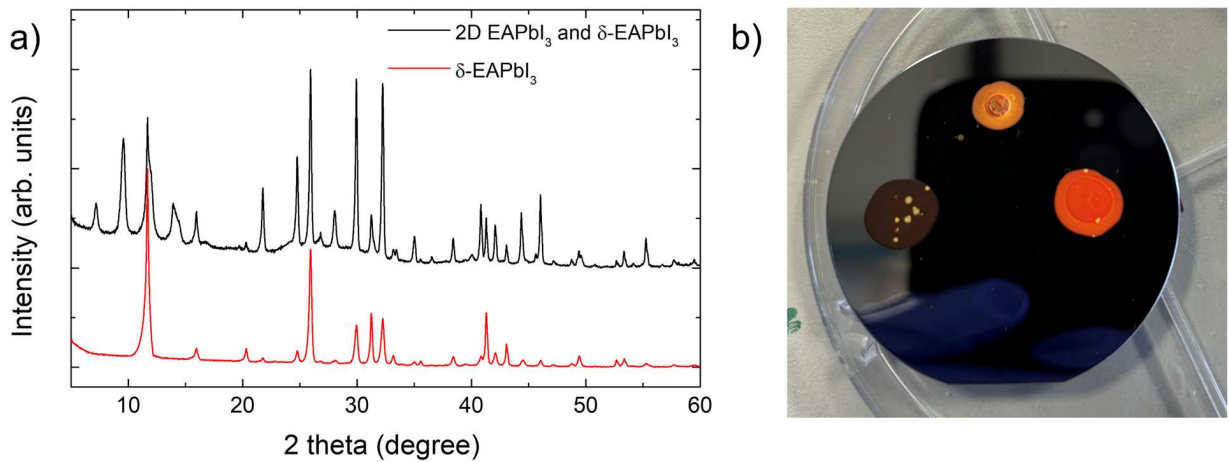

**Figure S10.** (a) Transformation of the EAPbI₃ perovskite phase to both 2D structures and the non-perovskite δ-EAPbI3 simultaneously. Repetitive peaks at the low angle indicate the formation of 2D structures. Further, peaks at 11° and 26° indicate the formation of δ-EAPbI₃ in the black line. The red line demonstrates pure δ-EAPbI₃. (b) Digital images of the drop cast films in the case of the transformation from EAPbI₃ perovskite phase to 2D and δ-EAPbI₃. The film's color was changed from black to orange. Also, it is possible to see some yellow spots related to δ-EAPbI₃ formation in the digital image.

## S2. EAPbI₃/PbI₂ epitaxial interface modeling

The epitaxial relations between EAPbI$_3$ and PbI$_2$ were identified using the Ogre Python library for the prediction of epitaxial interfaces.[2–4] For simplicity, we adopted a pseudocubic description for EAPbI$_3$, while for PbI$_2$ we selected the trigonal reference structure proposed in previous study.[5] Since the relative orientation of the two domains could be inferred from the HAADF-STEM images, we did not perform a full exploration of the possible epitaxial relations. Instead, we used Ogre to identify the 2D-supercells describing the two observed epitaxial interfaces (Figure S11) and to propose an optimized structure model thereof.

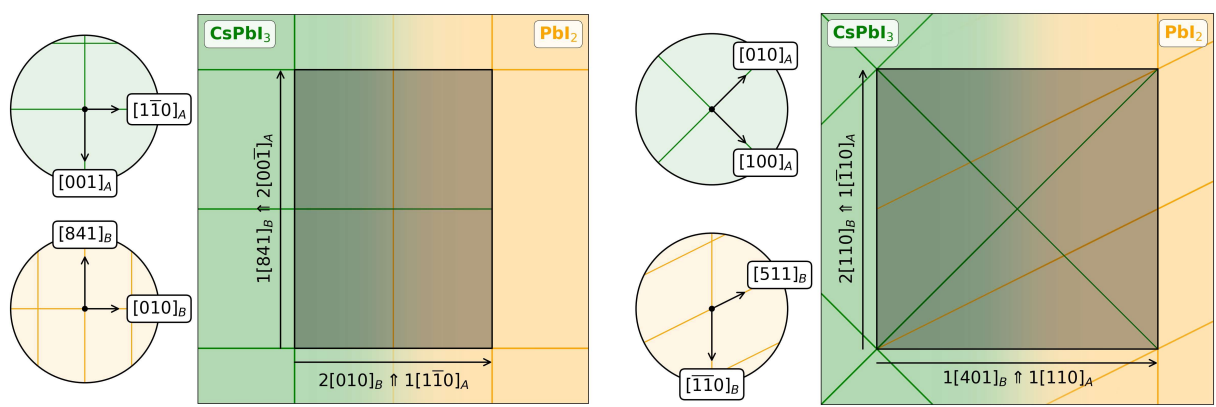

**Figure S11.** Interface supercells and relative orientation of the EAPbI$_3$ and PbI$_2$ lattices (in green and orange respectively) for the epitaxial interfaces shown in Figure 2e (left) and Figure 2f (right). The first interface occurs between the (110) plane of EAPbI$_3$ and the (10-8) plane of PbI$_2$, and is characterized by a supercell area of 116 Å$^2$ and 6 % strain. The second interface occurs between the (001) plane of EAPbI$_3$ and the (-114) plane of PbI$_2$, and is characterized by a supercell area of 82 Å$^2$ and 6 % strain. We note that this second epitaxial relation is equivalent to the one reported in Ref.[5] or the PbI$_2$/FAPbI$_3$ system. See Ref.[2] for details on the interpretation of these diagrams.

To perform the interface model optimization, we replaced $EA^+$ in the structure with $Cs^+$, because the version of Ogre we employed does not support structures involving organic cations. However, the lattice constant was kept to the 6.4 Å value we measured for $EAPbI_3$, and because the Ogre code adopts a fully classical electrostatic potential for energy evaluation, replacing one monovalent cation for another is expected to have limited impact on the outcome of simulations.

For both interfaces the structure proposed by Ogre was describing adequately the position of heavy atoms visible in the HAADF-STEM images reported in Figure 2. Notably, both models appear well connected, with the $I^-$ anions of $EAPbI_3$ bonding naturally with the $Pb^{2+}$ cations of $PbI_2$ so to complete their octahedral coordination environment.[2] Note that Ogre tends to overestimate the bonding distance at the interface due to the inability of its simplified electrostatic potential to fully capture interactions that are quantomechanical in their essence fully. For visualization purposes, in Figure 2 of the Main text we corrected such distance manually by taking as a reference the Pb-I bond length found in the two bulk materials (~3.2 Å). The raw output structures proposed by Ogre are attached as a part of the supplementary material.

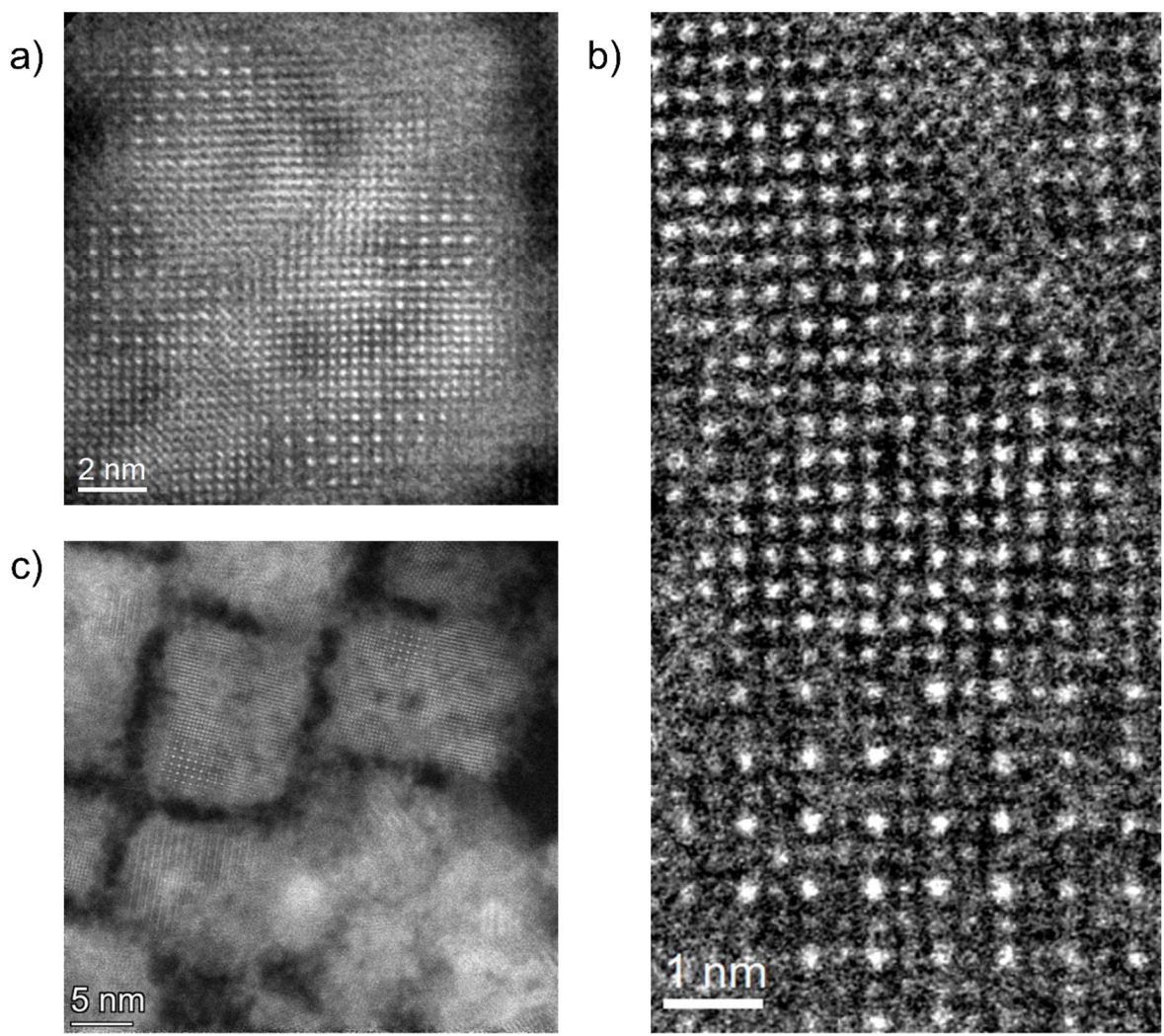

**Figure S12.** HAADF images of (a) $(EA_xCs_{1-x})PbI_3$ perovskite domains together with trigonal $PbI_2$ domain in the single NC. The formation of $(EA_xCs_{1-x})PbI_3/PbI_2$ heterostructures can be explained by the instability of the $EAPbI_3$ phase under an electron beam. (c) Lower magnification image of the $(EA_xCs_{1-x})PbI_3/PbI_2$ heterostructures. Figure 2b shows an atomic resolution HAADF image of the interface between $(EA_xCs_{1-x})PbI_3/PbI_2$ domains. The upper part of the images corresponds to the trigonal $PbI_2$ lattice at [-22-3] projection; it is connected to the $(EA_xCs_{1-x})PbI_3$ domain. Due to the very high beam sensitivity of the material, this analysis was made only qualitatively based on the intensity of the corresponding atomic columns. In the HAADF image the brightest columns

correspond to Pb/I due to the highest average Z number. The minimum contrast corresponds to higher EA containing ($EA_xCs_{1-x}$) columns. With an increase in Cs doping the contrast of (EA, Cs) columns is increased. The I columns in the (($EA_xCs_{1-x}$)$PbI_3$) have a contrast similar to the (EA, Cs) columns enriched in Cs, due to the little difference in Z numbers of I and Cs.

**Table 1.** Extracted lattice parameters from XRD patterns of the $EA_xFA_{(1-x)}PbI_3$ NCs. The corresponding composition (here indicating the fraction of $EA^+$ and $FA^+$ present in the alloy), was estimated by applying Vegard's law: $x = (d_x - d_B)/(d_A - d_B)$, where $d_x$, $d_A$, and $d_B$ are the measured lattice constants of the alloy and of the pure materials *A* and *B*, respectively.

| Material | Lattice constant (Å) | Composition ($EA_xFA_{(1-x)}PbI_3$) |
| --- | --- | --- |
| $EAPbI_3$ | 6.4308 | 1.00 |
| $EAPbI_3$ + 7 µL FA-oleate | 6.3980 | 0.61 |
| $EAPbI_3$ + 14 µL FA-oleate | 6.3919 | 0.54 |
| $FAPbI_3$ | 6.3460 | 0.00 |

**Table 2.** Extracted lattice parameters from XRD patterns of the $EA_xFA_{(1-x)}PbI_3$ NCs. The corresponding composition (here indicating the fraction of $EA^+$ and $FA^+$ present in the alloy) was estimated by applying Vegard's law: $x = (d_x - d_B)/(d_A - d_B)$, where $d_x$, $d_A$, and $d_B$ are the measured lattice constants of the alloy and of the pure materials *A* and *B*, respectively.

| Material | Lattice constant (Å) | Composition ($EA_xCs_{(1-x)}PbI_3$) |
|---|---|---|
| EAPbI$_3$ | 6.4308 | 1.00 |
| EAPbI$_3$ + 3.5 µL Cs-oleate | 6.3722 | 0.73 |
| EAPbI$_3$ + 7 µL Cs-oleate | 6.3481 | 0.62 |
| EAPbI$_3$ + 14 µL Cs-oleate | 6.2786 | 0.29 |
| CsPbI$_3$ | 6.2151 | 0 |

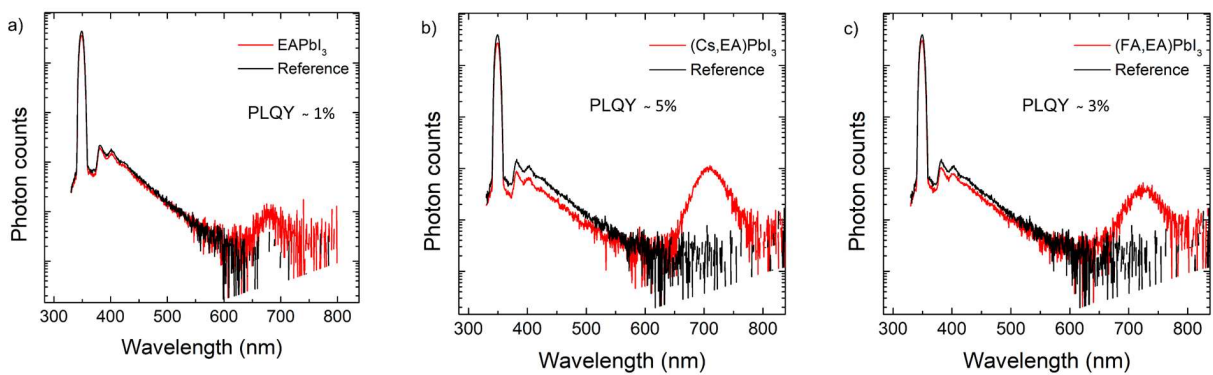

**Figure S13.** Spectra recorded for PLQY measurement of (a) EAPbI$_3$, (b) (Cs, EA)PbI$_3$, and (c) (FA, EA)PbI$_3$ perovskite NCs, respectively.

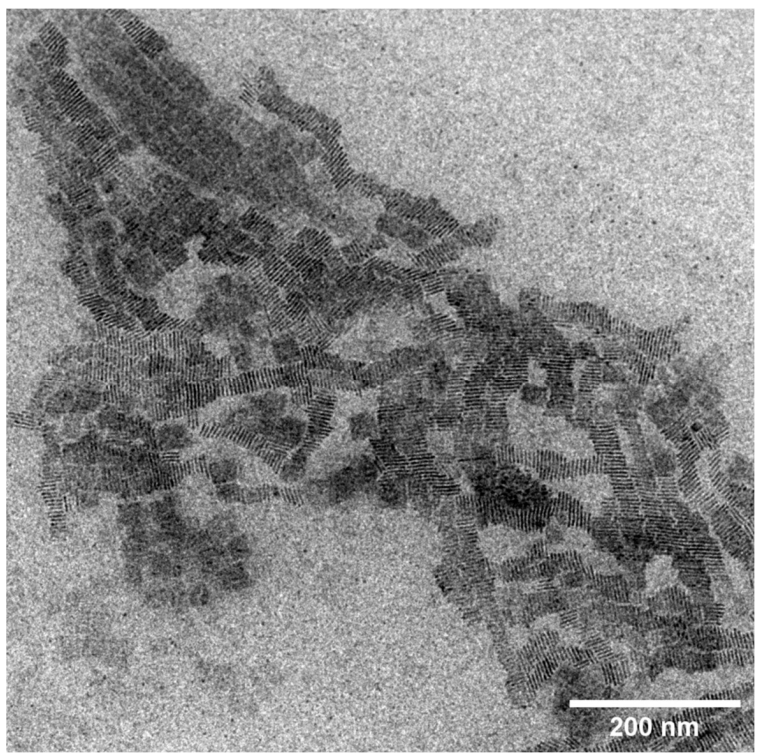

**Figure S14.** (a) Low magnification and TEM images of the 2 monolayers EAPbI$_3$ nanoplatelets.

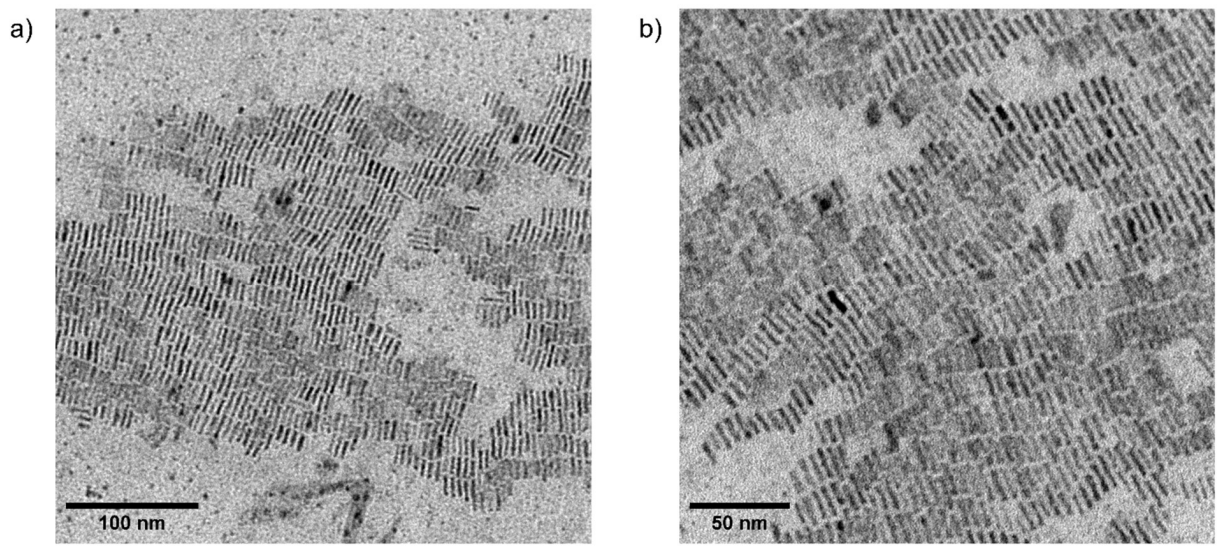

**Figure S15.** (a, b) Bright field TEM images of the mixed thickness EAPbI$_3$ nanoplatelets in different magnifications.

## S3. EAPbI₃ nanoplatelets analysis

The XRD pattern of EAPbI$_3$ nanoplatelets shown in Figure 4d demonstrates the characteristic features of multilayer diffraction, that is a secondary interference effect occurring in XRD patterns of highly ordered assemblies of colloidal nanocrystals. The phenomenon is described in detail in Refs.[6,7], but in short it consists in the formation of sharp fringes due to the constructive interference of radiation scattered by neighboring particles. For nanoplatelets, this XRD profile can be simulated, and in the best cases refined, using the code developed by us in Ref.[6].

Prior to the analysis, the pattern must be prepared by performing a manual background subtraction (Figure S16, top), followed by the subtraction of residual peaks that do not belong to the equally spaced series of multilayer interference fringes (Figure S16, bottom). In this specific case, two of the residual peaks (q = 1.0 Å$^{-1}$ and q = 2.0 Å$^{-1}$) belong to EAPbI$_3$ nanoplatelets that are positioned perpendicular to the silicon substrate, and therefore do not take part to multilayer interference, while the peak at 2.2 Å$^{-1}$ is a spurious reflection from the substrate. From the EAPbI$_3$ residual peaks we can extract the lateral lattice periodicity (and therefore Pb-Pb distance), that is 6.35 Å.

Following the background and spurious peak subtraction, the XRD pattern was analyzed using our code from Ref.[6] Although the quality of data was insufficient for a detailed refinement of the nanoplatelets structure, which would allow to precisely determine the position of all heavy atoms, the XRD profile is fully compatible with a thickness of 2 PbI$_6$ octahedra, corresponding to a nominal stoichiometry of (OLAM)$_2$EAPb$_2$I$_7$. The nanoplatelets stacking periodicity was estimated from the fit to 39.5 Å, while the disorder parameter is estimated to $\sigma_L$ = 0.8 Å (see Ref.[6] for a full description of the method and parameters).

Another relevant parameter extracted from the fit is the vertical distance of the two $Pb^{2+}$ ions layers found in the platelet structure. This was found to be 6.31 Å, which is similar to the horizontal Pb-Pb distance in the same platelets (see above) and much smaller compared to the Pb-Pb distance in $EAPbI_3$ cuboidal nanocrystals. We clarify that the intensity profile in Figure S16 appears different from both Figure S15 and Figure 4 of the Main Text (note the intensity ratio of peaks in the 0.5 – 1.25 Å$^{-1}$ and 1.5 – 2.25 Å$^{-1}$ range) because our data analysis script works by fitting the square module of the structure factor (shown here), and not the experimental intensity (shown in the main text). The two profiles can be interconverted via the Lorentz-Polarization correction factor.[8]

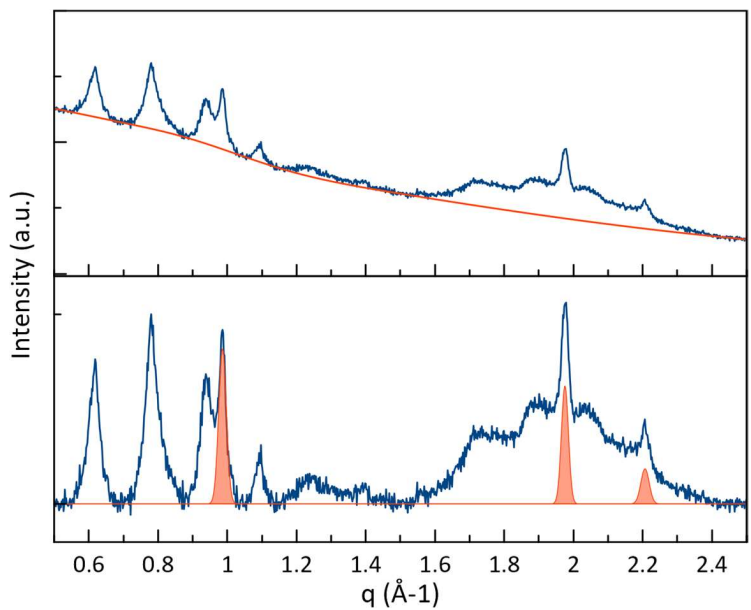

**Figure S16.** XRD pattern of EAPbI3 nanoplatelets plotted in *q*-scale (q = 4π sin(θ)/λ$_{X-ray}$), with background highlighted (top), and after background subtractions with residual peaks highlighted (bottom). manual background subtraction (top). The pattern resulting from the subtraction of both background and residual peaks has been used as an input for the multilayer diffraction analysis script from Ref.[6] (see Figure S16).

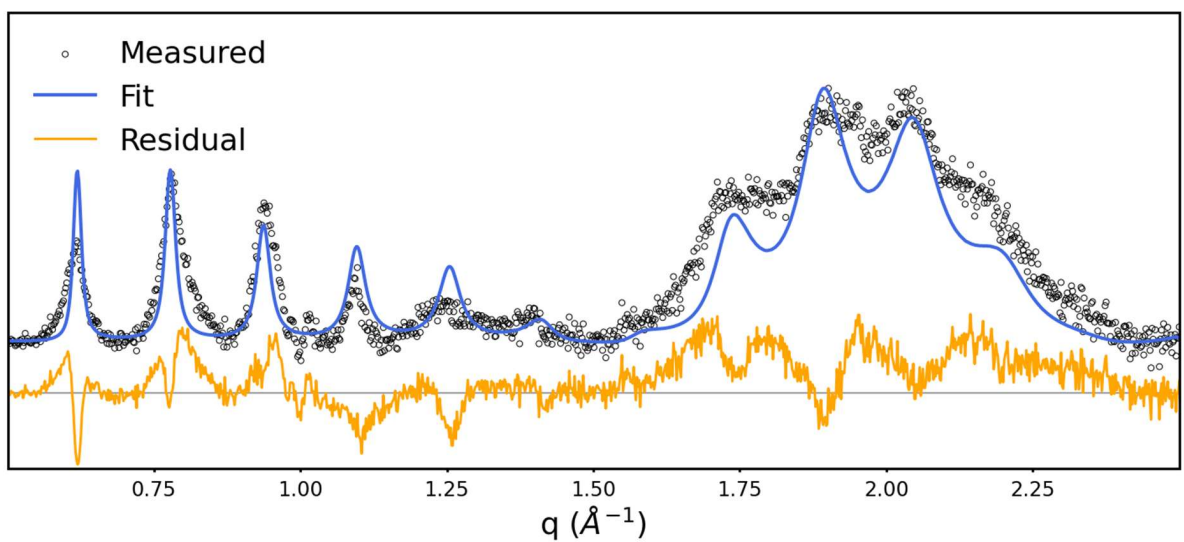

**Figure S17.** Multilayer diffraction simulation of the EAPbI$_3$ platelets XRD pattern. The high signal/background ratio does not allow a high quality refinement, but the simulation is sufficient to estimate the platelets stacking periodicity and vertical Pb-Pb distance with decent accuracy.

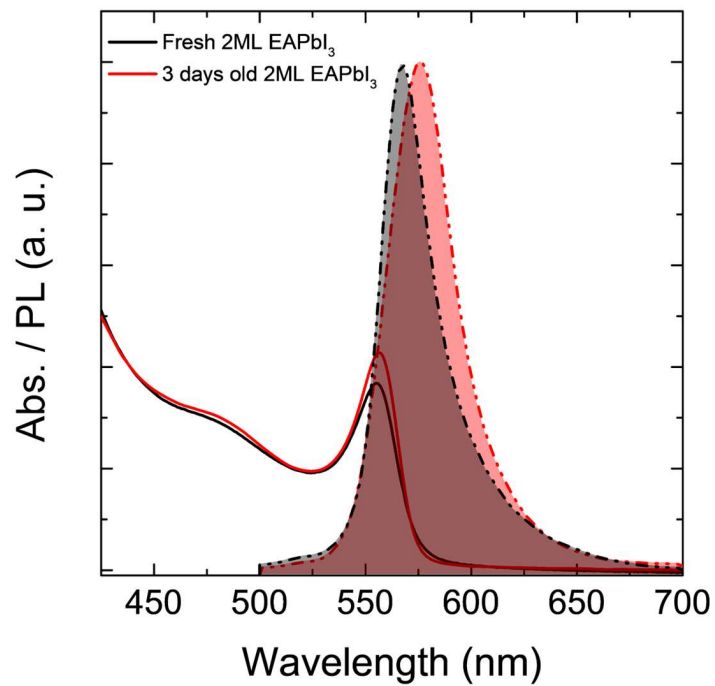

**Figure S18.** Optical absorbance and PL spectra of the 2 ML EAPbI$_3$ nanoplatelets as prepared (black) and after 3 days of aging (red) under ambient conditions in a closed vial.